\documentclass[prb,11pt]{revtex4-1}
\usepackage{amsmath}    
\usepackage{color}
\begin{document}%

\title{Massive electrodynamics for London's superconductivity and Josephson effect}
\author{A. I. Arbab }
\affiliation{Department of Physics, College of Science, Qassim University, P.O. Box 6644, 51452 Buraidah,  KSA}

\begin{abstract}
Massive electrodynamics for London's superconductivity and Josephson effect are derived. The propagation of massive boson inside a medium yields electric phenomena that are reflected in the Josephson effect. Critical force, magnetic field and temperature are found to be related to the critical current of the junction. The mass of the boson depends only on the critical current of the junction. The electromagnetic interaction between the Cooper pairs in the two sides of the superconductor in the josephson junction is mediated by a massive boson. The propagation of the electromagnetic waves mediated by the massive bosons gives rise to the electric properties of the Josephson junction. Of these properties are  a quantized resistance of Hall type corresponding to a non-quantized magnetic flux, and a quantized capacitance. A non zero magnetic flux encompassing a magnetic charge is found to arise despite the fact that it is not a priori assumed.
\end{abstract}

\maketitle

\section{\textcolor[rgb]{0.00,0.07,1.00}{Introduction}}

In Josephson effect, the tunneling of Cooper pairs across a superconducting junction, two superconductors sandwiched by a thin insulator, is found to give rise to a critical current without voltage. When a potential difference is applied the current oscillates. This is due to a phase difference associated with the pairs in the two superconductor \textcolor[rgb]{0.00,0.07,1.00}{\cite{jose}}. The current is considered to result from tunneling of the Cooper pairs across the junction. We treat in this work the Josephson effect as a quantum electrodynamics effect than rather a single quantum particle effect.

We show in this work that electromagnetic effects resulting from propagation of massive boson through the junction induce effects similar to the one observed in Josephson experiments. This reflects the matter wave associated with motion of massive photons. Maxwell's equations describing such interactions are obtained by introducing interactions with these photons (bosons). In these equations a current term that is proportional to the magnetic field is found to lead to this effect. No special concern is made for the superconductor. The only assumption is that the electric potential energy is equal to the rest mass energy of the boson, \emph{i.e.}, $q\varphi=mc^2$\,, where $m$ is the mass of the boson. Whether this case is satisfied for a superconductor or not is a matter of dispute.

The question whether this is a generic effect or not is  also a matter of question. Moreover, one may argue in favor that the Cooper pair is the boson that mediates the electromagnetic interaction inside the superconductor. Other phenomena that result from such an effect are also shown. This relies on the assumption that the magnetic flux is quantized inside the junction. Of these phenomena are the critical residence of the junction and the critical power that the junction can deliver. A critical applied magnetic field  is  found to be related to the critical current generated.

If the power delivered by the junction is that  of a black body radiation,  the critical current and magnetic field are found to  be related to the junction temperature. The mass of the boson is found to depend only on the critical current of the junction. The interrelation between quantum and electromagnetic effects are found to be responsible for these behaviors.

We study in Section II the massive bosons in the Maxwell's equations by adopting the Proca - Maxwell's equations where we derive the two London's equations.  In Section III we benefit from the idea of the minimal coupling applied to Dirac's equation to explore the interaction of the massive bosons with the electrons. This inscription leads to some quantum effect associated with electromagnetic interactions which finds an application in the Josephson junction.  We associate in this section several electromagnetic quantities we the due to the interaction of the massive bosons with the electron. We study in Section IV special set of Maxwell's equations that has important applications in the field of condensed matter. We end this work with some concluding remarks.

\section{\textcolor[rgb]{0.00,0.07,1.00}{Massive bosonic field  in Maxwell's theory}}

Maxwell's equations describing the interaction of massless (non-interacting)  photons are given by \textcolor[rgb]{0.00,0.07,1.00}{\cite{griff}}
\begin{equation}\label{1}
\vec{\nabla}\cdot\vec{E}=\frac{\rho}{\varepsilon_0}\,,\qquad\qquad\qquad\qquad \vec{\nabla}\cdot\vec{B}=0
\end{equation}
\begin{equation}\label{1}
\vec{\nabla}\times\vec{E}+\frac{\partial \vec{B}}{\partial
t}=0\,,\qquad\qquad \vec{\nabla}\times\vec{B}-\frac{1}{c^2}\frac{\partial
\vec{E}}{\partial t}=\mu_0\vec{J}\,.
\end{equation}
When photons are considered to be interacting, the electromagnetic interactions are described by the Maxwell-Proca equations \textcolor[rgb]{0.00,0.07,1.00}{\cite{proca}}, \emph{viz}.,
\begin{equation}\label{1}
\vec{\nabla}\cdot\vec{E}=\frac{\rho}{\varepsilon_0}-\mu^2\varphi\,,\qquad\qquad\qquad\qquad\qquad \vec{\nabla}\cdot\vec{B}=0\,,
\end{equation}
\begin{equation}\label{1}
\vec{\nabla}\times\vec{E}+\frac{\partial \vec{B}}{\partial
t}=0\,,
\qquad\qquad
\vec{\nabla}\times\vec{B}-\frac{1}{c^2}\frac{\partial
\vec{E}}{\partial t}=\mu_0\vec{J}-\mu^2\vec{A}\,,
\end{equation}
where $\mu$ is related to the mass of the boson, $\vec{A}$ and $\varphi$ are the vector and scalar potentials. Recall that the electric  and magnetic fields are defined by
$$\vec{E}=-\vec{\nabla}\varphi-\frac{\partial\vec{A}}{\partial t}\,,\qquad\qquad\vec{B}=\vec{\nabla}\times\vec{A}\,.$$
While Maxwell's equations are Lorentz and gauge invariant, the Maxwell-Proca's equations are not gauge invariant. This merit makes eqs.(3) and (4) to be less predictive. Here the electromagnetic interactions are mediated by a massive spin-1 vector boson, $\vec{A}$. This vector field satisfies the Klein-Gordon equation. The vector $\vec{A}$ is proportional to the current density associated with  the motion of the massive boson, and the scalar $\varphi$ is proportional to the charge density associated with the massive boson. The electromagnetic interactions inside a superconductor is found to be of short range, and therefore are governed by the Maxwell-Proca equations.

Manipulations of eqs. (3) and (4) yield the following equations
\begin{equation}\label{1}
\frac{1}{c^2}\frac{\partial^2\vec{E}}{\partial t^2}-\nabla^2\vec{E}+\mu^2\vec{E}=-\mu_0\left(c^2\vec{\nabla}\rho+\frac{\partial\vec{J}}{\partial t}\right)\,,\qquad
\frac{1}{c^2}\frac{\partial^2\vec{B}}{\partial t^2}-\nabla^2\vec{B}+\mu^2\vec{B}=\mu_0\vec{\nabla}\times\vec{J}\,,
\end{equation}
\begin{equation}\label{1}
\frac{1}{c^2}\frac{\partial^2\vec{A}}{\partial t^2}-\nabla^2\vec{A}+\mu^2\vec{A}=\mu_0\vec{J}-\vec{\nabla}\left(\vec{\nabla}\cdot\vec{A}+\frac{1}{c^2}\frac{\partial\varphi}{\partial t}\right)\,,\qquad \frac{1}{c^2}\frac{\partial^2\varphi}{\partial t^2}-\nabla^2\varphi+\mu^2\varphi=\frac{\rho}{\varepsilon_0}\,,
\end{equation}
and
\begin{equation}\label{1}
\vec{\nabla}\cdot\vec{J}+\frac{\partial\rho}{\partial t}=\frac{\mu^2}{\mu_0}\,\left(\vec{\nabla}\cdot\vec{A}+\frac{1}{c^2}\frac{\partial\varphi}{\partial t}\right)\,.
\end{equation}
In vacuum, where $\rho=0$ and $\vec{J}=0$, eqs.(5), (6) and (7) reveal that $\vec{A}$, $\vec{E}$ and $\vec{B}$ satisfy the Klein-Gordon equation. Moreover, eq.(7) shows that the Lorenz gauge condition and the charge conservation are connected.

Instead of the conventional vacuum, one can define a rather general vacuum as the one which satisfies the following equations
\begin{equation}\label{1}
\vec{\nabla}\rho+\frac{1}{c^2}\,\frac{\partial\vec{J}}{\partial t}=0\,,\qquad\qquad\vec{\nabla}\times\vec{J}=0\,, \qquad \qquad
\vec{\nabla}\cdot\vec{J}+\frac{\partial\rho}{\partial t}=0\,.
\end{equation}
In this vacuum, Maxwell's equations reveal that the electromagnetic field travels at the speed of light.
This set of equations is found to be of great importance \textcolor[rgb]{0.00,0.07,1.00}{\cite{contin}}. Owing to eq.(8),  the vacuum is not necessarily empty but could retain some sort of matter satisfying eq.(8). It is evident from eq.(8) that an incompressible fluid, \emph{i.e}., $\rho=\rm const.$ and $\vec{J}=\rho\,\vec{v}$, is stationary and has a vanishing vorticity, $\vec{\omega}=\vec{\nabla}\times\vec{v}=0$\,.

For a conducting medium, one can apply eq.(8) in Maxwell's equations with $\vec{J}=\sigma\,\vec{E}$. This yields
$$
\frac{1}{c^2}\,\frac{\partial^2\rho}{\partial t^2}=\nabla^2\rho\,,\qquad\qquad \frac{1}{c^2}\,\frac{\partial^2\vec{E}}{\partial t^2}=\nabla^2\vec{E}\,,\qquad\qquad \frac{\partial\vec{B}}{\partial t}=0\,,\qquad\qquad\nabla^2\vec{B}=0\,.\!\!\!\!\!\qquad\qquad (8a)
$$
Thus, for a non-conventional conducting vacuum, the electric field and the charge density satisfy a wave equation travelling at the speed of light, whereas the magnetic field is time-independent satisfying Laplace equation.

In London's theory of superconductivity, the electric current, charge density, electric and magnetic fields are related by \textcolor[rgb]{0.00,0.07,1.00}{\cite{lond}}
\begin{equation}\label{1}
c^2\vec{\nabla}\rho+\frac{\partial\vec{J}}{\partial t}=\frac{\mu^2}{\mu_0}\,\vec{E}\,,
\end{equation}
\begin{equation}\label{1}
\vec{\nabla}\times\vec{J}=-\frac{\mu^2}{\mu_0}\vec{B}\,,
\end{equation}
where  $\frac{n_se^2}{m}=\frac{\mu^2}{\mu_0}$, where $n_s$ and $m$ are  the number density and mass of the superelectrons \textcolor[rgb]{0.00,0.07,1.00}{\cite{lond}}. If we now apply eq.(9) and (10) in eqs.(5) and (6), we see that the electric and magnetic fields of London satisfy Maxwell-Proca equations (the Klein-Gordon equation) with massive boson whose mass is equal to $\sqrt{2}\,\mu$. A glance at eqs.(6), (9) and (10) reveals that one should have the following relations
\begin{equation}\label{1}
\vec{J}=-\frac{\mu^2}{\mu_0}\,\vec{A}\,,\qquad\qquad \rho=-\varepsilon_0\mu^2\varphi\,,
\end{equation}
so that $\vec{A}$ and $\varphi$ satisfy the same Klein-Gordon equation as that of $\vec{E}$ and $\vec{B}$, with mass equals to $\sqrt{2}\,\mu$. Therefore, the system of equations, eqs.(9) - (11), comprises the London's equations of superconductivity.

Equation (7) can be seen as defining a total current that is conserved. This is evident if we write eq.(7) as
\begin{equation}\label{1}
\vec{\nabla}\cdot\vec{J}_T+\frac{\partial\rho_T}{\partial t}=0\,,\qquad \vec{J}_T=\vec{J}-\frac{\mu^2}{\mu_0}\,\vec{A}\,,\qquad \rho_T=\rho-\varepsilon_0\mu^2\varphi\,.
\end{equation}
The current and charge densities defined in eq.(11) can be seen as that due to the massive boson. The current $\vec{J}_T$ is always conserved no matter whether Lorenz gauge condition is satisfied or not. This is not unexpected since the bosons interact with the electromagnetic field.

The energy conservation equation can be obtained from eqs.(3) - (4) and the charge conservation equation in eq.(7) that reads
\begin{equation}\label{1}
\vec{\nabla}\cdot\vec{S}_{\rm eff.}+ \frac{\partial u_{\rm eff.}}{\!\!\!\partial t}=-\vec{E}\cdot\vec{J}\,,
\end{equation}
where
\begin{equation}\label{1}
\vec{S}_{\rm eff.}=\frac{\vec{E}\times\vec{B}+\mu^2\varphi\,\vec{J}}{\mu_0}\,,\qquad u_{\rm eff.}=\frac{1}{2}\,\varepsilon_0(E^2+\mu^2\varphi^2)+\frac{1}{2\mu_0}\,(B^2+\mu^2A^2)\,.
\end{equation}
Interestingly, eq.(12) states that the scalar and vector potentials are of physical significance, since they have energy and momentum. This clearly shows that the bosonic field propagates along the direction of the current. This conforms with our recent work related to the analogy between matter and electromagnetic waves \textcolor[rgb]{0.00,0.07,1.00}{\cite{epl}}. They can be attributed to the material nature of the bosonic field. Remarkably, the bosonic current, $\propto\,\vec{A}$, is non-dissipative, since it doesn't appear in the right - hand side of the energy conservation equation above. The boson field seems to be appearing along the surface.

\section{\textcolor[rgb]{0.00,0.07,1.00}{The Josephson effect}}

In a recent work we have found that one can introduce an interaction between photons by allowing their mass to be non-zero but in a different way to that of Proca. This results in the modified-Maxwell's equations incorporating interactions terms \textcolor[rgb]{0.00,0.07,1.00}{\cite{arbab}}. These read
\begin{equation}\label{1}
\vec{\nabla}\cdot\vec{E}=\frac{\rho}{\varepsilon_0}-\frac{qc}{\hbar}\vec{A}_g\cdot\vec{B}-\frac{q}{\hbar}\varphi_g\Lambda\,,
\end{equation}
\begin{equation}\label{1}
\vec{\nabla}\times\vec{E}+\frac{\partial \vec{B}}{\partial
t}=-\frac{q}{\hbar c}\left(\,\varphi_g \,\vec{E}+c^2\vec{A}_g\times\vec{B}\right)+c\vec{\nabla}\Lambda\,,
\end{equation}
\begin{equation}
\vec{\nabla}\times\vec{B}-\frac{1}{c^2}\frac{\partial
\vec{E}}{\partial t}=\mu_0\vec{J}-\frac{q}{\hbar c}\left(\,\varphi_g\, \vec{B}-\vec{A}_g\times\vec{E}\right)-\frac{q}{\hbar}\vec{A}_g\Lambda\,,
\end{equation}
and
\begin{equation}
\vec{\nabla}\cdot\vec{B}=\frac{q}{\hbar c}\vec{A}_g\cdot\vec{E}+\frac{1}{c}\frac{\partial \Lambda}{\partial t}\,,
\end{equation}
where $\vec{A}_g$ and $\varphi_g$ are the potentials of the background field (bosons). When $\Lambda=0$, the above equations suggest effective electric charge and current densities of the system as $\rho_{e}=\rho-\frac{q}{\mu_0\hbar c}\vec{A}_g\cdot\vec{B}$ and $\vec{J}_{e}=\vec{J}-\frac{\mu_0q}{\hbar c}\,\varphi_g\, \vec{B}-\frac{\mu_0q}{\hbar c}\,\vec{A}_g\times\vec{E}$, and magnetic charge and current densities as $\rho_m=\frac{q}{\hbar c}\vec{A}_g\cdot\vec{E}$, and $\vec{J}_m=\frac{q\varphi_g }{\hbar c}\,\vec{E}+\frac{qc}{\hbar }\,\vec{A}_g\times\vec{B}$. Thus, these effective charges and currents result from the interaction of the electromagnetic field with the background field. This interaction carries a quantum signature. It could also be seen as resulting from the deformation of the space  created by the background field.

These equations reduce to Wilczek axion electrodynamics \textcolor[rgb]{0.00,0.07,1.00}{\cite{wilc}}. They have a quantum character reflecting the massive nature of the bosons. They can also be seen as representing Maxwell's equations in a background field characterised by $\vec{A}_g$ and $\varphi_g$. Notice that the in the case of source-free system, the above modified Maxwell's equations are invariant under the duality transformation, \emph{viz.}, $\vec{E}\rightarrow c\vec{B}$ and $c\vec{B}\rightarrow-c\vec{E}$. A similar set of equations is also proposed by Majern\'{i}k to describe the interaction of light in presence of some scalar field \textcolor[rgb]{0.00,0.07,1.00}{\cite{majern}}

Let us now consider the case $\vec{A}_g=0$ and $\varphi_g=\rm const.$ in eqs.(15) - (18). This yields
\begin{equation}
\vec{\nabla}\cdot\vec{E}=\frac{\rho}{\varepsilon_0}\,,\qquad\qquad \vec{\nabla}\cdot\vec{B}=0\,,
\end{equation}
\begin{equation}
\vec{\nabla}\times\vec{E}+\frac{\partial \vec{B}}{\partial
t}=-\frac{q\,\varphi_g}{\hbar c}\,\vec{E}\,,
\end{equation}
\begin{equation}
\vec{\nabla}\times\vec{B}-\frac{1}{c^2}\frac{\partial
\vec{E}}{\partial t}=\mu_0\vec{J}-\frac{q\,\varphi_g}{\hbar c}\, \vec{B}\,.
\end{equation}
Unlike Maxwell-Proca equations, the energy  due to the above system is the same as that for Maxwell's equations, \emph{i.e}., the energy conservation equation is independent of the potential $\varphi_g$.
The wave equations associated with  eqs.(19) - (21) are given by
\begin{equation}\label{1}
\frac{1}{c^2}\frac{\partial^2\vec{E}}{\partial t^2}-\nabla^2\vec{E}+\beta^2\vec{E}=-2\beta\,\vec{\nabla}\times\vec{E}-\frac{1}{\varepsilon_0}\left(\vec{\nabla}\rho+\frac{1}{c^2}\frac{\partial\vec{J}}{\partial t}\right)\,,
\end{equation}
\begin{equation}
\frac{1}{c^2}\frac{\partial^2\vec{B}}{\partial t^2}-\nabla^2\vec{B}+\beta^2\vec{B}=-2\beta\,\vec{\nabla}\times\vec{B}+\mu_0\left(\vec{\nabla}\times\vec{J}+\beta\,\vec{J}\right),
\end{equation}
where $\beta=\frac{q\varphi_g}{\hbar\,c}$ is constant. When $\varphi_g=const$, we have $\rho=0$, and $\vec{\nabla}\cdot\vec{J}=0$, as can be seen  by taking the divergence of eq.(20) and using eq.(19). Hence, eq.(22) can be written as
\begin{equation}
\frac{1}{c^2}\frac{\partial^2\vec{E}}{\partial t^2}-\nabla^2\vec{E}+\beta^2\vec{E}=-2\beta\,\vec{\nabla}\times\vec{E}-\mu_0\,\frac{\partial\vec{J}}{\partial t}\,.
\end{equation}
Let us now take the dot product of the surface element vector of eq.(20) and integrate the resulting equation to obtain the potential difference (voltage)
\begin{equation}
V=-\frac{\partial\phi_B}{\partial t}-\frac{q\,\varphi_g}{\hbar\,c\varepsilon_0}\,Q\,\,,\qquad\qquad \varepsilon_0\int \vec{E}\cdot d\vec{s}=Q\,.
\end{equation}
The potential difference in the second term in eq.(25) can be written as
\begin{equation}
V_c=\frac{q\,\varphi_g}{\hbar\,\varepsilon_0c}\,Q\,.
\end{equation}
This suggests a quantum capacitance which is associated with the charge $Q$ that can be defined as
\begin{equation}
C=\frac{\varepsilon_0c}{q\,\varphi_g}\,\hbar\,\,.
\end{equation}
This could represent the capacitance resulting from the accumulation of charged bosons residing on the two sides of the junction.

If the right-hand sides of eqs.(23) and (24) vanish then $\vec{E}$ and $\vec{B}$ will satisfy the Klein-Gordon equation. In this case one has
\begin{equation}
-2\beta\,\vec{\nabla}\times\vec{B}+\mu_0\left(\vec{\nabla}\times\vec{J}+\beta\,\vec{J}\right)=0\,,
\end{equation}
and
\begin{equation}
2\beta\,\vec{\nabla}\times\vec{E}+\mu_0\,\frac{\partial\vec{J}}{\partial t}=0\,.
\end{equation}
Taking the dot product of eq.(29) with surface element vector and integrating the resulting equation yield a quantum potential difference
\begin{equation}
V_b=-\frac{\partial }{\partial t}\,\left(\frac{\mu_0c}{2q\varphi_g}\, I\right)\hbar\,,
\end{equation}
from the changing flux, $\phi_b$,  that induces a current
\begin{equation}
I=\frac{2q\varphi_g}{\mu_0c\hbar}\, \phi_b\,,
\end{equation}
which gives rise to a quantum inductance
\begin{equation}
L_b=\frac{\mu_0c}{2q\varphi_g}\,\,\hbar\,,
\end{equation}
and a quantum resistance
\begin{equation}
R_b=\frac{\varphi_g}{I}=\frac{\mu_0c}{2q\,\phi_b}\,\hbar\,,
\end{equation}
when the magnetic flux $\phi_b$ is not quantized. One can associate $\varphi_g$ to the transverse potential. A transient quantum voltage due to the flow of massive boson can be associated with eq.(30) that is developed during a time $\Delta t$. This can be defined as
\begin{equation}
V_t=\left(\frac{\mu_0c}{2q}\frac{\hbar}{q\varphi_g}\right)\,I^2\,, \qquad{\rm or}\qquad \qquad V_t=\left(\frac{\mu_0}{2q}\,\frac{\,\,\hbar}{mc}\right)\,I^2\,,
\end{equation}
where we have set $\Delta t=q/I$ and $q\varphi_g=mc^2$. Observe that the current -voltage relationship is not like that one in the ordinary Ohm's relationship. It is an equation of a parabola. It could also mean that the resistance increases with current.

Furthermore, applying eq.(20) in eq.(29) and rearranging eq.(28) yield two equations analogous to that of eqs.(9) and (10), with $\rho=0$, \emph{viz.},
\begin{equation}
\frac{\partial\vec{J}_{\rm eff.}}{\partial t}=\frac{2\beta^2}{\mu_0}\,\vec{E}\,,
\end{equation}
and
\begin{equation}
\vec{\nabla}\times \vec{J}_{\rm eff.}=-\beta\,\vec{J}\,,
\end{equation}
where $ \vec{J}_{\rm eff.}=\vec{J}-\frac{2\beta}{\mu_0}\,\vec{B}$ and $\mu^2=2\beta^2$.

The extra currents in eqs.(20) and (21) are non-dissipative currents.
The first term in the right - hand side of eq.(21) indicates that for a conductor the current density ($ \vec{J}=\sigma\,\vec{E}$) is along the direction of the electric field, the last term suggests a current density perpendicular to the electric field, and the second term  suggests a current density of the form $\vec{J}_q=\frac{q\varphi_g}{\mu_0c\hbar}\,\vec{B}$, or a current
\begin{equation}
I_q=\frac{q\varphi_g\phi_B}{\mu_0c\hbar}\,\,,
\end{equation}
along the direction of the magnetic field, where $\phi_B$, is the magnetic flux. This relation suggests a magnetic conductivity, $\sigma_m$ given by $\sigma_m=\frac{q\varphi_g}{\mu_0c\hbar}$. This is dubbed  the \emph{Chiral Magnetic Effect} (CME) and $\sigma_m$ is called the chiral magnetic conductivity \textcolor[rgb]{0.00,0.07,1.00}{\cite{cme1, cme2}}. This effect is known to be of topological origin. In these models $\varphi_g$ represents  the chiral chemical potential. We would like here to analyze the implications of such a current.

If we now assume in eq.(37) that the flux is quantized, \emph{i.e.}, $\phi_B=\frac{\hbar}{q}\,n$, then eq.(36) leads to a characteristic resistance, $R_q=\frac{\varphi_g}{I_q}=\frac{\mu_0c}{n}$\,, where $n$ is an integer. Recall that the impedance of free space is defined by $Z_0=\mu_0c$. Hence, one can write,
\begin{equation}
R_q=\frac{Z_0\,}{n}\,.
\end{equation}
This urges us to relate $R_q$ to the impedance that massive photons experience inside a medium. Now if the flux is not quantized, then eq.(37) will give a resistance
\begin{equation}
R=\left(\frac{\mu_0c}{q\phi_B}\right)\,\hbar\,\,,
\end{equation}
which reveals that it has a quantum character. Furthermore, eq.(39) dictates that when $\phi_B=\mu_0c\,q$ the resistance will be equal to the quantum Hall resistance, $R=\frac{\hbar}{q^2}$. It is interesting to see that in Gauss's law the electric flux ($\phi_E=\frac{Q}{\varepsilon_0}$) is proportional to the total charge ($Q$) enclosed by the Gaussian surface. In a similar manner, one can assume that the magnetic flux is proportional to the magnetic charge enclosed. In this case, the magnetic charge associated with this flux should be defined as $q_m=cq$ so  that $\phi_B=\mu_0q_m$. It is surprising that eq.(19) does not presume a priori the existence of magnetic charge, however.

Taking the dot product of eq.(21) with surface element and integrating yield
\begin{equation}
\oint_C\vec{B}\cdot d\vec{\ell}=\mu_0\left(i+\frac{\partial Q}{\partial t}-\frac{q^2}{\hbar}\,\varphi_g\right)\,,
\end{equation}
along a closed contour $C$, where $Q$ is the total charge enclosed by the contour. The third current (negative) in the parentheses can be expressed as $I_h=\frac{q^2}{\hbar}\,\varphi_g$ which gives rise to a Hall-like resistance $R=\frac{\hbar}{q^2}$, where $\phi_B=\mu_0cq$.

We see that the electromagnetic wave mediated by massive boson propagates through the insulator in the junction without dissipation. The maximum value of $R_q$ is when $n=1$. Moreover,  an inductance
\begin{equation}
L_q=\frac{\mu_0\hbar c}{q\varphi_g}\,,
\end{equation}
can be associated with the  current $I_q$. If we let $q\varphi_g=mc^2$, one has $L_q=\frac{\mu_0\hbar}{mc}$. It is shown by Beck that the current in the Josephson junction peaks at this value \textcolor[rgb]{0.00,0.07,1.00}{\cite{beck}}. This can be written as $L_q=\mu_0\lambda_c$\,, where $\lambda_c=\hbar/(mc)$ is the Compton wavelength of the massive photon. This coincides with our quantum inductance that is found to be associated with massive photon \textcolor[rgb]{0.00,0.07,1.00}{\cite{massive}}. If we now set $q=2e$ and $q\varphi_g=mc^2$, the current $I_q=\frac{mc}{2e\mu_0}\,n$, which can be written as
\begin{equation}
I_q=\frac{\hbar}{2eL_q}\,n\,.
\end{equation}
Interestingly, this current is quantized. However, in the Josephson  junction the critical current ($I_0$) depends on the applied magnetic field and the junction dimensions \textcolor[rgb]{0.00,0.07,1.00}{\cite{jose}}. It is found to be related to the critical inductance ($L_0$)  by the relation, $L_0=\hbar/(2eI_0)$. This corresponds to $L_q$ when $n=1$. The critical current is a quantum mechanical manifestation of the phase difference of the Cooper pairs wavefunctions in the two sides of the junction. This current is found to be in the range $1\mu A-1nA$. It sets a mass limit
\begin{equation}
m=\frac{\mu_0(2e)}{c}\,I_q\,,
\end{equation}
for the boson (photon) in the range $\sim 10^{-38}-10^{-41}\rm kg$ $(1meV-1\mu eV)$, if it were responsible for the Josephson effect. This may suggest that axion and massive photon are but the same object. It was assumed by Beck that axion induced-current can manifest itself as a supercurrent in Josephson junction \textcolor[rgb]{0.00,0.07,1.00}{\cite{beck}}. It is suggestive to call the above mass an \emph{electromagnetic  mass}, since it depends solely on electromagnetic quantities.
One can associate a characteristic length dimension ($\lambda_0=\frac{\hbar}{mc}$), \emph{e.g}., width  of the junction,  with the above equation so that the resulting critical current reads
\begin{equation}
I_q=\frac{\hbar}{2e\,\mu_0\lambda_0}\,\frac{1}{n}\,.
\end{equation}
While the current $I_0$ is attributed to quantum tunneling of Cooper pairs, the current $I_q$ is attributed to the transmission of the massive field (photon). The energy transmitted by  the massive field ($\Delta E$) through the junction can be considered to be related to the time during which it acts via the Heisenberg's uncertainty relation. Hence, one can write, $\Delta E\,\Delta t=\hbar\,,$ where $\Delta E=E_q$ is the Josephson coupling energy, and $\tau\equiv\Delta t=q/I_q$ is the time during which the current is transmitted. This yields, setting $q=2e$,
\begin{equation}
E_q=\frac{\hbar I_q}{2e}\,,
\end{equation}
that coincides with the exact relation obtained by Josephson \textcolor[rgb]{0.00,0.07,1.00}{\cite{jose}}. The agreement between the two approaches is due to the fact electromagnetic and inertial properties of a particle are related. This is evident from the fact that Maxwell's equations emerge from Dirac equation in the higher representation (quaternionic) \textcolor[rgb]{0.00,0.07,1.00}{\cite{quatern}}. Thus, the use of any approach will lead to the same result. For instance, the Cooper pair uses its inertial (quantum) property to tunnel through the junction, but uses its electromagnetic nature in its subsequent motion.

Using the above equation, one can associate a critical power with the Josephson junction as
\begin{equation}
P_q=\mu_0c\,I_q^2\,,\qquad\qquad P_q=Z_0I_q^2\,.
\end{equation}
This power is due to the massive photon transporting the junction. And since $I_q$ is generally very small, the radiated power is exceedingly small.
It seems that the current flows through the insulator as if it were vacuum. So what really flows is the boson that is a mediator. Thus, the current that flows inside the insulator is the boson current.
A current in the range  of $1\mu A$ sets an output power of the order $10^{-10}\rm W.$ This value is confirmed by Minami \emph{et al}. who found that such a power  emerges from a single Josephson junction \textcolor[rgb]{0.00,0.07,1.00}{\cite{power}}. The power in eq.(46) can also be obtained if one defines this power as $P_q=mc^2/\tau$ and uses eq.(43). If the above power is emitted as a black-body, then
\begin{equation}
I_q=\sqrt{\frac{A\sigma_e}{\mu_0c}}\,\,T_q^2\,,\qquad\qquad mc^2=2e\,\sqrt{\mu_0c\sigma_eA}\,\,T_q^2\,,\qquad\qquad \varphi=\sqrt{\mu_0c\sigma_eA}\,\,T_q^2\,.
\end{equation}
where $T_q$ is the gas (black body) absolute temperature, $A$ is the area of the junction, and $\sigma_e$ is the Stefan's constant. Furthermore, one can relate the critical  magnetic field and the temperature, $T_q$, as
\begin{equation}
B_q=\sqrt{\frac{\mu_0\sigma_e}{c}}\,\, T_q^2\,.
\end{equation}
Therefore, the critical magnetic  field to be applied on the junction area $A$ to induce a critical current $I_q$ is given by
\begin{equation}
B_q=\frac{\mu_0}{\sqrt{A}}\,\, I_q\,.
\end{equation}
If we further assume that the above power is a consequence of Larmor power due to the massive photon, then its critical acceleration will be
\begin{equation}
a_q=\sqrt{\frac{3\pi}{2}}\,\frac{c}{e}\,I_q\,,
\end{equation}
inside the junction. For instance a critical current of $I_q\sim 1\mu A$ will give rise to an acceleration of $a_q\sim 4\times 10^{21}\rm m\,s^{-2}$. The critical force associated with this acceleration is
\begin{equation}
F_q=\sqrt{\frac{3\pi}{2}}\,\frac{mc}{e}\,I_q\,,\qquad {\rm or}\qquad\qquad F_q=\sqrt{6\pi}\,\mu_0I_q^2\,.
\end{equation}
This is similar to Ampere's tensile force that will appear in a wire when a high current is applied on it \textcolor[rgb]{0.00,0.07,1.00}{\cite{massive}}. We have shown that this occurs if the force acting on the massive photon inside the junction is a viscous (drag) force \textcolor[rgb]{0.00,0.07,1.00}{\cite{massive}}. A typical value of this force amounts to $F_q\sim 10^{-10}\rm N$. Can we connect this force to some short-range attractive electric force between the Cooper pairs in the two sides of the junction? If this is of Coulomb type, then the Cooper pairs should be at a distance of a few nanometers.

In Josephson junction the critical current is related to the superconducting gap energy at $T=0$\,,\emph{ i.e}., $\Delta(0)$, by the Ambegaokar-Baratof relation \textcolor[rgb]{0.00,0.07,1.00}{\cite{ambag}}
\begin{equation}
I_0=\frac{\pi\,\Delta(0)}{2eR_N}\,\,,
\end{equation}
where $R_N$  is the normal state resistance of the junction. One can therefore, write
\begin{equation}
\frac{R_N}{h/(2e)^2}=\frac{\Delta(0)}{2E_J}\,\,.
\end{equation}
Hence, if $E_J=\frac{1}{2}\,\Delta(0)$, then $R_N=h/(2e)^2$\,, which is the Hall resistance of the junction. In this case the coupling energy can be measured and the above relations can be verified.

\section{\textcolor[rgb]{0.00,0.07,1.00}{Special modified Maxwell's equation}}
Let us consider the case when $\Lambda\ne0$ but $\vec{A}_g=0$ and $\varphi_g=0$. Applying this case in Eqs.(15) - (18) yields
\begin{equation}\label{1}
\vec{\nabla}\cdot\vec{E}=\frac{\rho}{\varepsilon_0}\,,
\end{equation}
\begin{equation}\label{1}
\vec{\nabla}\times\vec{E}=-\frac{\partial \vec{B}}{\partial
t}+c\vec{\nabla}\Lambda\,,
\end{equation}
\begin{equation}
\vec{\nabla}\times\vec{B}=\frac{1}{c^2}\frac{\partial
\vec{E}}{\partial t}+\mu_0\vec{J}\,,
\end{equation}
and
\begin{equation}
\vec{\nabla}\cdot\vec{B}=\frac{1}{c}\frac{\partial \Lambda}{\partial t}\,.
\end{equation}
It is to be noted that the wave equations for the electromagnetic fields are not altered by the presence of $\Lambda$ in Eqs.(55) and (57). It is remarkable for $\vec{A}_g\ne0$ and $\varphi_g\ne0$, that Eqs.(15) and (17) suggest  additional electric charge and electric current densities of the form
\begin{equation}\label{1}
\vec{J}_\Lambda=-\frac{q\Lambda}{\mu_0\hbar}\, \vec{A}_g\,,\qquad\qquad \rho_\Lambda=-\frac{\varepsilon_0\varphi_g\Lambda}{\hbar}\,,
\end{equation}
having a structure of  London's current of superconductivity but with quantum character \textcolor[rgb]{0.00,0.07,1.00}{\cite{lond}}. In London'd theory $\vec{J}=-\frac{nq^2}{m}\,\vec{A}$. This implies that $\Lambda=\frac{\mu_0q\hbar}{m}\,n$. And since $n$ is a function of temperature, we anticipate $\Lambda$ to be a function of temperature too.

Equations (15) - (18) admit  magnetic charge and current densities of the form
\begin{equation}\label{1}
\rho_m=\frac{1}{c}\frac{\partial \Lambda}{\partial t}\,,\qquad\qquad \vec{J}_m=-c\vec{\nabla}\Lambda\,,
\end{equation}
respectively. As can be seen from Eq.(58), the magnetic charge is conserved, as far as $\Lambda$ satisfies the wave equation. Moreover, if $\Lambda$ is attributed to some thermal scalar field, then the magnetic charge and current densities can alternatively  represent some thermal current and charge densities that are  associated with temperature gradients.

The energy conservation equation associated with Eqs.(54) - (57) is
\begin{equation}\label{1}
\vec{\nabla}\cdot\vec{S}_\Lambda+\frac{\partial u_\Lambda}{\partial
t}=-\vec{E}\cdot\vec{J}\,,
\end{equation}
where
\begin{equation}\label{1}
\vec{S}_\Lambda=\frac{\vec{E}\times\vec{B}-c\Lambda\vec{B}}{\mu_0}\,,\qquad\qquad u_\Lambda=\frac{\varepsilon_0}{2}E^2+\frac{B^2}{2\mu_0}+\frac{\Lambda^2}{2\mu_0}\,.
\end{equation}
Equation (61) shows that some additional energy flows along the magnetic field direction besides the transverse one flowing along a direction perpendicular to  the electric and magnetic fields directions. Equation (55) and (57) show that $\Lambda$ satisfies the wave equation traveling at the speed of light in vacuum. There exists an interesting case when the electric field vanishes, \emph{i.e}.,  $\vec{E}=0$, so that Eq.(60) yields a dissipation-less energy conservation equation, \emph{viz}.,
\begin{equation}\label{1}
\vec{\nabla}\cdot\vec{S}_\Lambda+\frac{\partial u_\Lambda}{\partial
t}=0\,,
\end{equation}
where
\begin{equation}\label{1}
\vec{S}_\Lambda=-\mu_0^{-1}c\Lambda\,\vec{B}\,,\qquad\qquad u_\Lambda=\frac{B^2}{2\mu_0}+\frac{\Lambda^2}{2\mu_0}\,.
\end{equation}
A physical situation corresponding to this particular case should be seek.  It is  magnetic wave flowing along the magnetic field direction carrying magnetic energy. It is interesting to see from Eqs.(15) - (18) that the presence of $\Lambda$ induces electric charge, magnetic charge, electric current, and magnetic current densities. This amounts to say that $\Lambda$ is some kind of background (internal) field prevailing the space that is coupled to the magnetic field.

The momentum conservation equation associated with the vanishing electric field can be obtained from Eqs.(54) - (57),
\begin{equation}
\frac{\partial \vec{B}}{\partial
t}=c\vec{\nabla}\Lambda\,,
\qquad\qquad
\vec{\nabla}\times\vec{B}=\mu_0\vec{J}\,,\qquad\qquad
\vec{\nabla}\cdot\vec{B}=\frac{1}{c}\frac{\partial \Lambda}{\partial t}\,,
\end{equation}
 as
\begin{equation}\label{1}
\frac{1}{c^2}\frac{\partial S_i}{\partial t}+\partial_j T_{ij}=(\vec{J}\times\vec{B})_i\,,
\end{equation}
where
\begin{equation}\label{1}
T_{ij}=\mu_0^{-1}(B_iB_j+\frac{1}{2}(\Lambda^2-B^2)\,\delta_{ij})\,,\qquad\qquad S_i=-\mu_0^{-1}c\Lambda B_i\,.
\end{equation}
The term $\vec{S}/c^2$ can be seen as the momentum density, and $T_{ij}$ as the magnetic stress tensor. Note that $\vec{S}$ can be positive or negative depending on the sign of $\Lambda$. Thus, the scalar field $\Lambda$ can flow along or against the magnetic field direction.

\section{\textcolor[rgb]{0.00,0.07,1.00}{Concluding remarks}}

By introducing an interaction between matter field and  electromagnetic field, we obtained a set of modified Maxwell's equations. Consequently, a massive boson is found to mediate the electromagnetic interactions. This boson field satisfies the Klein-Gordon equation. The London's theory of superconductivity is found to be a particular case of the Maxwell-Proca equations. Ampere's equation reveals a current density that is parallel to the magnetic field. This current gives rise to electric effect as that of Josephson effect. A critical  current is found to be associated with the mass of the boson. This mass depends solely on the critical current of the junction. Moreover,  critical magnetic field, power, inductance, resistance,  temperature, force are found to be connected with this critical current. These quantities are a result of the propagation of the massive bosons and the material nature associated with them. The derived critical quantities await experimental verification. The scalar field, $\Lambda$ is shown to lead to interesting manifestations in Maxwell's equations. It induces additional electric, magnetic charge and current densities when coupled to the background scalar and vector potentials.


\begin{thebibliography}{}


\bibitem{jose}
 Josephson B. D., 1962 \emph{Phys. Lett.} 1 251.

\bibitem{griff}
Griffiths D., 1999 \emph{Introduction to Electrodynamics}  (Prentice-Hall).

\bibitem{proca}
Proca A., 1936 \emph{J. Phys. Radium Ser.} VII 7 347.

\bibitem{contin}
Arbab A. I. Widatallah H. M., 2010 \emph{Chin. Pys. Lett.} 27  084703.

\bibitem{lond}
London F. and London H., 1935)\emph{Proc. Roy. Soc. A} 149 866.

\bibitem{epl}
 Arbab A I., 2011  \emph{Europhys. Lett.}  94 50005.

\bibitem{arbab}
Arbab A. I., 2017, to appear in Int. J. Mod. Phys. A.


\bibitem{wilc}
Wilczek F., 1987 \emph{Phys. Rev. Lett.} 58, 1799;  1978 \emph{Phys. Rev. Lett.} 40 279.


\bibitem{majern}
Majernik V., 1982, Astrophys. Space Sci., 84 191.

\bibitem{cme1}
Fukushima K. ,  Kharzeev D. E.,  Warringa H. J., 2008 \emph{Phys. Rev. D} 78 074033.

\bibitem{cme2}
D'Hoker E, Goldstone J., 1985, \emph{Phys. Lett. B} 158 429.

\bibitem{beck}
 Christian B., 2013 \emph{Phys. Rev. Lett.} 111 231801.


\bibitem{massive}
Arbab A I., 2014, \emph{Progress In Electromagnetics Research M} 34 153.

\bibitem{quatern}
Arbab A I., 2017 \emph{Maxwellian quantum mechanics} Optik  136, 382.

\bibitem{power}
 Minami H,  \emph{et al.}, 2012 \emph{Journal of Physics: Conference Series} 400 022072.

\bibitem{ambag}
Ambegaokar V and  Baratoff A., 1963 \emph{Phys. Rev. Lett.} 10 486.

\end{thebibliography}
\end{document}